\documentclass[12pt, preprint,aps,prc,showpacs,superscriptaddress,tightenlines,nofootinbib]{revtex4}
\usepackage{epsf}
\usepackage{amsmath}

\def\beq{\begin{equation}}
\def\eeq{\end{equation}}
\def\bea{\begin{eqnarray}}
\def\eea{\end{eqnarray}}
\def\beqa{\begin{equation}\begin{array}{l}}
\def\eeqa{\end{array}\end{equation}}
\def\eqlab#1{\label{eq:#1}}
\def\figlab#1{\label{fig:#1}}

\def\eref#1{(\ref{eq:#1})}
\def\Eqref#1{Eq.~(\ref{eq:#1})}

\def\Figref#1{Fig.~\ref{fig:#1}}

\def\slap{p \hspace{-2mm} \slash}

\def\half{\mbox{\small{$\frac{1}{2}$}}}

\def\quarter{\mbox{\small{$\frac{1}{4}$}}}

\def\sixth{\mbox{\small{$\frac{1}{6}$}}}

\def\barr{\left(\begin{array}{c}}
\def\earr{\end{array}\right)}
\def\bmat{\left(\begin{array}{cc}}
\def\emat{\end{array}\right)}

\def\al{\alpha}
\def\be{\beta}
\def\ga{\gamma} 
\def\de{\delta} \def\De{\Delta}

\def\veps{\varepsilon}  \def\eps{\epsilon}

\def\la{\lambda} \def\La{{\Lambda}}

\def\si{\sigma} 
\def\th{\theta}  
\def\w{\omega}

\def\pa{\partial}

\def\pa{\partial}

\def\nn{\nonumber}

\def\mathscr{\mathcal}

\def\arctg{{\rm arctg}}

\def\3d{3-D}

\def\amm{a.m.m.}
\def\rhs{r.h.s.}
\def\lhs{l.h.s.}

\begin{document}
\preprint{WM-05-111}

\title{Sum Rules for Magnetic Moments and Polarizabilities in QED and Chiral Effective-Field Theory }

\author{Barry R.\ Holstein}
\email{holstein@physics.umas.edu} \affiliation{Theory Group, JLab,
12000 Jefferson Ave, Newport News, VA 23606}
\affiliation{Department of Physics--LGRT, University of
Massachusetts, Amherst, MA 01003}

\author{Vladimir Pascalutsa}
\email{vlad@jlab.org}
\affiliation{Theory Group, JLab, 12000 Jefferson Ave, Newport News, VA 23606}
\affiliation{Department of Physics, College of William \& Mary, Williamsburg, VA
 23188}

\author{Marc Vanderhaeghen}
\email{marcvdh@jlab.org}
\affiliation{Theory Group, JLab, 12000 Jefferson Ave, Newport News, VA 23606}
\affiliation{Department of Physics, College of William \& Mary, Williamsburg, VA
 23188}
\date{\today}

\begin{abstract}
We elaborate on a recently proposed extension of the Gerasimov-Drell-Hearn
(GDH) sum rule which is achieved by taking derivatives
with respect to the anomalous magnetic moment.  The new sum rule features
a {\it linear} relation between the anomalous magnetic moment and
the dispersion integral over a cross-section quantity.
We find some analogy of the linearized form of the GDH sum rule with
the `sideways dispersion relations'.
As an example, we apply the linear sum rule to reproduce
the famous Schwinger's correction to the magnetic moment in QED
from a tree-level cross-section calculation and outline the procedure
for computing the two-loop correction from a one-loop cross-section
calculation. The polarizabilities of the electron in QED are considered as well
by using the other forward-Compton-scattering sum rules.
We also employ the  sum rules
to study the magnetic moment and polarizabilities of the nucleon
in a relativistic chiral EFT framework. In particular we investigate the
chiral extrapolation of these quantities.
\end{abstract}

\pacs{13.60.Fz - Elastic and Compton scattering.
14.20.Dh - Proton and neutrons.
25.20.Dc - Photon absorption and scattering}%
\maketitle
\thispagestyle{empty}

\section{Introduction}
The Gerasimov, Drell, Hearn (GDH) sum rule (SR)~\cite{GDH}, which relates a
system's anomalous magnetic moment to a weighted integral over a
combination of doubly polarized photoabsorption cross sections,
has received a good deal of attention in recent years.  Impressive
experimental programs to measure these photoabsorption cross-sections for the
nucleon have recently been carried out at ELSA and MAMI (for a
review see Ref.~\cite{Drechsel:2004ki}).  Such measurements
provide an empirical test of the GDH SR, and can be used to
generate phenomenological estimates of electromagnetic
polarizabilities via related SRs, as will be discussed below.
The GDH SR is particularly interesting because both its left- and
right-hand-side can be reliably determined, thus providing a
useful verification of the fundamental principles (such as
unitarity and analyticity) which go into its derivation.  At the
present time the proton sum rule is satisfied within the
experimental precision, while the case is still out for
the neutron.  However, it is not the purpose of this paper to
discuss these experimental aspects of the SR, but rather to see
what can be learned on the theoretical side.

Recently, we have shown~\cite{Pascalutsa:2004ga} that by taking
derivatives of the GDH sum rule with respect to the anomalous
magnetic moment one can obtain a new set of sum rule-like
relations with intriguing properties.  In particular, this
procedure provided a sum rule involving the anomalous magnetic
moment {\it linearly} rather than quadratically, which allows for
the derivation of quantities such as the Schwinger moment in a
much simpler fashion than via the usual GDH method.  In this
paper, we shall further examine these forms and apply them to the
nucleon magnetic moment.

After a lightning review of the GDH and related sum rules
(Sect.~II), we derive the modified versions (Sect.~III) and
demonstrate that some of these relations have the form of
so-called {\it sideways} dispersion relations (Sect.~IV). Then we
consider applications to the nucleon in the context of chiral
perturbation theory and show how the new sum rules allow an
elementary calculation (to one loop) of quantities such as
magnetic moments (Sect.~V) and polarizabilities (Sect.~VI) to all
orders in the heavy baryon expansion.  The chiral behavior of the
nucleon magnetic moments and polarizabilities is addressed in
Sect.~VII. Returning to QED in Sect.~VIII, we demonstrate how the
new sum rules can be applied to the two-loop calculation of the
anomalous magnetic moment in a straightforward fashion. In the
final section, we summarize our findings and suggest prospects for
future work.

\section{Compton-scattering sum rules in QED}
The forward-scattering amplitude describing the elastic scattering
of a photon on a target with spin $s$ (real Compton scattering) is
characterized by $2s+1$ scalar functions which depend on a single
kinematic variable, {\it e.g.}, the photon energy $\w$. In the
low-energy limit each of these functions corresponds to an
electromagnetic moment---charge, magnetic dipole, electric
quadrupole, {\it etc.}---of the target. In the case of a spin-1/2
target, the forward Compton amplitude is generally written as
\begin{eqnarray}
\label{DDeq2.2.2} T(\w) = {\vec{\veps}\,'}^*\cdot\vec
\veps\,f(\w)+ i\,\vec
\sigma\cdot({\vec{\veps}\,'}^*\times\vec\veps)\,g(\w)\, ,
\end{eqnarray}
where $\vec \veps$, $\vec \veps\, '$ is the polarization vector of
the incident and scattered photon, respectively, while $\vec
\sigma$ are the Pauli matrices representing the dependence on the
target spin. The {\it
crossing symmetry} of the Compton amplitude of
Eq.~(\ref{DDeq2.2.2}) means invariance under
$\varepsilon'\leftrightarrow\varepsilon$, $\w\leftrightarrow-\w$, which
obviously leads to $f(\omega)$ being an even and
$g(\omega)$ being an odd function of the energy---$f(\w) =
f(-\w)$, $g(\w) =-g(-\w)$.
The two scalar functions $f(\omega),\,
g(\omega)$ admit the following low-energy expansion,
\begin{subequations}
\eqlab{let}
\begin{eqnarray}
f(\w) & = & -\frac{e^2}{4\pi M} + (\alpha_E+\beta_M)\,\w^2+ {\mathcal{O}}(\w^4) \ , \label{DDeq2.2.12} \\
g(\w) & = & -\frac{e^2\kappa^2}{8\pi M^2}\,\w +
\gamma_{0}\w^3 + {\mathcal{O}}(\w^5) \ , \label{DDeq2.2.13}
\end{eqnarray}
\end{subequations}
and hence, in the low-energy limit, are given in terms of the
target's charge $e$ and anomalous magnetic moment (a.m.m.)
$\kappa$. The next-to-leading order terms are given in terms of
the nucleon electric ($\al_E$), magnetic ($\be_M$), and forward
spin ($\ga_0$) polarizabilities.

In order to derive sum rules (SRs) for these quantities one
assumes the scattering amplitude is an {\it analytic} function of
$\w$ everywhere but the real axis~\footnote{Resonance poles may
occur but lie on the second Riemann sheet.}, which allows writing
the real parts of the functions  $f(\w)$ and $g(\w)$ as a {\it
dispersion integral} involving their corresponding imaginary
parts. The latter, on the other hand, can be related to
combinations of doubly polarized photoabsorption cross-sections
via the {\it optical theorem},
\begin{subequations}
\begin{eqnarray}
\mbox{Im}\ f(\w) & = & \frac{\w}{8\pi}
\left[\sigma_{1/2}(\w)+\sigma_{3/2}(\w)\right] \,, \\
\mbox{Im}\ g(\w) & = & \frac{\w}{8\pi}
\left[\sigma_{1/2}(\w)-\sigma_{3/2}(\w)\right] \,,
\label{DDeq2.2.6}
\end{eqnarray}
\end{subequations}
where $\sigma_{\la}$ is the doubly-polarized total cross-section
of the photoabsorption processes, with $\lambda$ specifying the
total helicity of the initial  system.  Averaging over the
polarization of initial particles gives the total unpolarized
cross-section,
 $\sigma_T=\half ( \sigma_{1/2}+\sigma_{3/2})$.

After these steps one arrives at the results (see, {\it e.g.},
\cite{Drechsel:2002ar} for more details): \bea \label{eq:drf}
f(\w) & = & f(0) \,+\, \frac{\w^2}{2\pi^2}\,
\int_{0}^{\infty}\frac{\sigma_T(\w')}
{\w'^2-\w^2 - i\eps} \, d\w'\ , \\
\label{eq:drg}
 g(\w) &=& -\frac{\w}{4\pi^2}\,
\int_{0}^{\infty}\frac{\De\sigma(\w')} {\w'^2-\w^2 -
i\eps}\,\w'\,d\w'\ , \eea with $\De\sigma\equiv
\sigma_{3/2}-\sigma_{1/2}$, and where the sum rule for the
unpolarized forward amplitude $f(\omega)$ has been once-subtracted
to guarantee convergence. These relations can then be expanded in
energy to obtain the SRs for the different static properties
introduced in \Eqref{let}. In this way we obtain the Baldin
SR~\cite{Bal60,Lap63}:
\begin{eqnarray}
\label{eq:baldinsr}
\alpha_E + \beta_M = \frac{1}{2\pi^2}\,
\int_{0}^{\infty}\,\frac{\sigma_T(\w)}{\w^2}
\,d\w \ ,
\end{eqnarray}
the GDH SR:
\begin{eqnarray}
\label{eq:gdhsr}
\frac{e^2\kappa^2}{2M^2}=\frac{1}{\pi}
\int_{0}^{\infty}\,\frac{\De\sigma(\w)}{\w}\,d\w \, ,
\end{eqnarray}
a SR for the forward spin polarizability:
\begin{eqnarray}
\label{eq:gamma0sr}
\gamma_0= \,-\,\frac{1}{4\pi^2}\,\int_{0}^{\infty}\,
\frac{\De\sigma(\w)}
{\w^3}\,d\w\ .
\end{eqnarray}
and, in principle, one could continue in order to isolate higher
order moments\cite{mainz}.

Let us now see how these sum rules are of use in field theory.
We first consider the case of the electron in
QED.
To lowest order in the fine-structure
constant, $\al_{em}=e^2/4\pi$, the photoabsorption process is
given by the tree-level Compton scattering. The tree-level helicity
amplitudes for this process are well known~\cite{okla}
\begin{equation}
\begin{array}{l}
T(++;++)={-8\pi\al_{em}\over
(M^2-s)^2(M^2-u)}((M^2-s)^2+M^2t)(M^4-su)^{1\over
2}\\
T(--;++)={-8\pi\al_{em}\over
(M^2-s)^2(M^2-u)}(-M^2t)^{3\over 2}, \\
T(-+;++)={8\pi\al_{em}\over
(M^2-s)^2(M^2-u)}(M^4-su)(-M^2t)^{1\over
2}, \\
T(++;+-)={8\pi\al_{em}\over
(M^2-s)^2(M^2-u)}M^2t(M^4-su)^{1\over 2}, \\
T(+-;+-)={-8\pi\al_{em}\over (M^2-s)^2(M^2-u)}(M^4-su)^{3\over
2}, \\
T(-+;+-)={-8\pi\al_{em}\over
(M^2-s)^2(M^2-u)}(-Ms)(-t)^{3\over 2},
\end{array}
\end{equation}
where $M$ is the electron mass; $s$, $t$ and $u$ are the Mandelstam variables.
 Using these results the
unpolarized and double-polarized cross sections can be
determined as:
\begin{eqnarray}
\sigma_{T}^{(2)}(\omega)
&=&{2\pi\al_{em}^2\over M^2}\left[{1+x\over
x^3}\left({2x(1+x)\over
1+2x}-\log(1+2x)\right)\right. \nonumber\\
&+&\left.{1\over 2x}\log(1+2x)-{1+3x\over
(1+2x)^2}\right]+{\cal O}(\al_{em}^3),
\label{eq:tot}
\end{eqnarray}
and
\begin{equation}
\De\sigma(\omega)=-{2\pi\al_{em}^2\over M^2x}\left[(1+{1\over
x})\log(1+2x)-2\left(1+{x^2\over (1+2x)^2
}\right)\right] +{\cal O}(\al_{em}^3),
\label{eq:gdhqed}
\end{equation}
where we have defined $x=\omega/M$. Substituting the latter
expression into the \rhs\ of the GDH SR, \Eqref{gdhsr}, one can
easily see that the integral vanishes exactly, which is required
because otherwise the sum rule would lead to a nonsense
result---the electron \amm\ would receive contributions of order
$\sqrt{\al_{em}}$.

At next order, ${\cal O}(\al_{em}^3)$, the \lhs\ of the GDH SR
receives a nonzero contribution in the form of the Schwinger
correction: $\kappa=\al_{em}/2\pi$.  In order to check that the
same result is obtained on the \rhs\ of the sum rule is quite a
formidable task since at this order one must know the Compton
scattering amplitude to one loop, as well as account for the
pair-production channel. Nevertheless, the calculation of the
relevant helicity amplitudes was carried out more than three
decades ago by Milton, Tsai, and deRaad and application to the GDH
SR has  relatively recently been performed by Dicus and
Vega~\cite{DiV01}, who verified (numerically) that the GDH SR
holds at ${\cal O}(\al_{em}^3)$ in QED.

Recently we have found a much simpler method by which to verify
the GDH SR at this order. This technique is briefly described in
the next section, while in the rest of the present  section we
examine the sum rules for polarizabilities,
Eqs.~(\ref{eq:baldinsr}) and (\ref{eq:gamma0sr}).  At first
glance, such sum rules do not appear to be of much utility, since
when evaluated in the case of QED, the static polarizabilities
obtained therefrom {\it diverge}.  However, this is not a problem,
as pointed out by Llanta and Tarrach~\cite{pol}, who emphasized
that one {\it can} determine well defined asymptotic forms in the
limit as $\omega$ approaches zero.  Thus, for example we can
perform the integration for the spin even/odd amplitudes and
determine that
\begin{equation}
f(\omega)-f(0)\simeq {\al_{em}^2\over \pi M}
\left[ {8\over 3}x^2\log 2x
+{11x^2\over 18}+{266\over 15}x^4\log 2x+{1799\over
450}x^4+\ldots \right] ,
\end{equation}
and
\begin{equation}
g(\omega)-g(0)\simeq {\al_{em}^2\over 2\pi M^2} \left[ {20\over
3}x^2\log 2x+{37\over 9}x^2+{896\over 15}x^4\log 2x+{29\cdot
128\over 225}x^4+\ldots \right].
\end{equation}
One can now use these expressions to define {\it quasi-static}
polarizabilities, where by this we mean that these quantities
contain both a constant term and a term which behaves as
$\log\omega$---it is this latter piece that is responsible for the
divergence as the static limit is taken.  Thus, for example, we
may define quasi-static values
\begin{equation}
\al_E^{q-s}+\beta_M^{q-s}={\al_{em}^2\over \pi M^3} \left[{8\over 3}\log 2x +{11\over 18} \right],
\end{equation}
for the sum of electric and magnetic polarizabilities and
\begin{equation}
\gamma_0^{q-s}={\al_{em}^2\over 2\pi M^4} \left[{20\over 3}\log
2x+{37\over 9} \right],
\end{equation}
for the forward spin polarizability.  In this way we can also
generate generalized sum rules for quadrupole and higher
polarizabilities via Eqs. \eref{drf} and
\eref{drg}~\cite{mainz}.

It is interesting to note that it is possible to determine the
{\it nonanalytic} component of these quasi-static moments in a
simpler fashion---by using only the the {\it low energy} expansion
of the cross sections.  That is, while the integrals
\begin{equation}
\int_0^\Lambda {dy y^n\over y^2-x^2},
\end{equation}
are $\Lambda$-dependent if $n$ is an even integer, in the case
that $n$ is odd there exists a $\Lambda$-independent logarithm
\begin{equation}
\int_0^\Lambda {dy y^{2\ell+1}\over y^2-x^2}=-x^{2\ell}\log x+...
\end{equation}
which then determines the logarithmic component of the
quasi-static polarizabilities.  That is, since
\begin{equation}
\sigma_{tot}(\omega)={2\pi\al_{em}^2\over M^2}\left[{4\over
3}-{8\over 3}{\omega\over M}+{104\over 15}{\omega^2\over
M^2}-{266\over 15}{\omega^3\over M^3}+\ldots\right],
\end{equation}
and
\begin{equation}
\sigma_{3\over 2}(\omega)-\sigma_{1\over 2}
(\omega)={2\pi\al_{em}^2\over M^2}\left[{4\over 3}{\omega\over
M}-{20\over 3}{\omega^2\over M^2 }+{108\over 5}{\omega^3\over
M^3}+{896\over 15}{\omega^4\over M^4 }+\ldots\right],
\end{equation}
we require that the nonanalytic piece of the polarizabilities have
the form
\begin{eqnarray}
&&\alpha_E^{q-s}+\beta_M^{q-s}={\al_{em}^2\over \pi M^2}
\left[{8\over 3}\log{\omega\over M}+... \right] \nonumber\\
&&\gamma_0^{q-s}={\al_{em}^2\over 2\pi M^4}\left[{20\over
3}\log{\omega\over M}+\ldots\right],
\end{eqnarray}
in agreement with the exact forms given above.  Likewise higher
order forms can be determined.

We see then that the use of dispersive techniques in QED allows a
straightforward extraction of information about polarizabilities
and about the anomalous magnetic moment.  As expected, the latter
is complete agreement with the result obtained by conventional
means, while the former {\it requires} such dispersive methods,
since the corresponding static quantities are divergent.

\section{Derivatives of the GDH sum rule}

We now review the derivation of the new form of sum rule.  We
begin by introducing a `classical' (or 'trial') value of the
electron \amm, $\kappa_0$. At the Lagrangian level this amounts to
the introduction of a Pauli term for the spin-1/2 field~: \beq
{\cal L}_{\mbox{\small Pauli}}= \frac{i\kappa_0}{4M} \, \bar
\psi\, \si_{\mu\nu}\, \psi\, F^{\mu\nu}\,, \eeq where $F^{\mu
\nu}$ is the electromagnetic field tensor and $\si_{\mu\nu}=(i/2)
[\ga_\mu,\ga_\nu]$ is the usual Dirac tensor operator.  At the end
of our calculation, we will set $\kappa_0$ to zero, but for now
the total value of the \amm\ is $\kappa=\kappa_0 +\delta \kappa$,
with $\de\kappa$ denoting the loop contribution.  (It is important
to note that both $\de\kappa$ and the cross-section become
explicitly dependent on $\kappa_0$.) We then start taking
derivatives of the GDH SR with respect to $\kappa_0$, which is
subsequently set to zero, so that the total \amm\ $\kappa$ returns
to its usual loop value.  We find
 \bea
(4\pi^2\al_{em}/M^2)\, \kappa \,\kappa' &=&
\int_{0}^\infty \!  \De\si'(\w)\, \frac{d\w}{\w}\,,\\
(4\pi^2\al_{em}/M^2)\, ( \kappa'^2+\kappa\,\kappa'')  &=&
\int_{0}^\infty \!  \De\si''(\w)\,\frac{d\w}{\w}\,,
\eea and so on.
To lowest order in $\al_{em}$ we find
 \beq
(4\pi^2\al_{em}/M^2)\, n\,\kappa^{(n-1)} = \int_{0}^\infty \!\!
\De\si^{(n)}(\w)\, \frac{d\w}{\w}\, , \label{eq:derivsr} \eeq
where $\kappa^{(n)}$ denotes the $n$th derivative of $\kappa$ with
respect to $\kappa_0$.  This allows in principle the computation
of $\kappa$ to order $\al_{em}^n$ by using the 1st to $n$th
derivatives of the cross-section computed to order
$\al_{em}^{n+1}$ to $\al_{em}^{2}$, respectively.

In particular, to lowest order we have the result~: \beq
\eqlab{linsr} \frac{4\pi^2\al_{em}}{M^2}\, \kappa  =
\int_{0}^\infty \! \left. \De\si'(\w)\right|_{\kappa_0=0}\,
\frac{d\w}{\w}\,. \eeq The striking feature of this sum rule is
the {\it linear} relation between the \amm\ and the (derivative of
the) photoabsorption cross section, in contrast to the GDH SR
where $\kappa$ appears quadratically, and although the
cross-section quantity $\De\si'(\omega)$ is not an observable, it
is very clear how it can be determined within a specific theory.
Thus, for example, the first derivative of the tree-level
cross-section with respect to $\kappa_0$, at $\kappa_0=0$, in QED
was worked out in~\cite{Pascalutsa:2004ga}: \beq
\left.\De\si'(\w)\right|_{\kappa_0=0} = \frac{2\pi\al_{em}^2}{M\w
}\left[ 6-{2M\w \over (M+2\w)^2} -\left(2+{3M\over
\w}\right)\,\ln\left(1+\frac{2\w}{M}\right) \right] .
\label{eq:gdh2} \eeq It is  not difficult to find then that
\begin{equation}
\eqlab{Schresult}
\frac{1}{\pi}\int\limits_0^\infty \!\left. \De\si'(\w)\right|_{\kappa_0=0}
\,\frac{d\omega}{\omega }
= \frac{2\al_{em}^2}{M^2} \,.
\end{equation}
Substituting this result in the {\it linearized} GDH SR,
\Eqref{linsr}, we obtain $\kappa= \al_{em}/2\pi$---Schwinger's
one-loop result. We emphasize that this result is reproduced here
by computing only a (derivative of the) {\it tree-level} Compton
scattering cross-section and then performing an integration over
energy. This is definitely much simpler than obtaining the
Schwinger result from the GDH SR directly~\cite{DiV01}, which
requires input at the one-loop level.  Along these lines, however,
one can facilitate the two- and more loop calculations. We
elaborate on this possibility in Sect.~VIII.


\section{Connection to sideways dispersion relations}

It is interesting to observe that, by changing the integration
variable to $s=M^2 + 2M\w$, the linearized GDH SR, \Eqref{linsr},
can be written as \beq \eqlab{sranddr} \kappa  =
\frac{M^2}{4\pi^2\al_{em}}\, \int\limits_{M^2}^\infty \! d s\,
\frac{\left. \De\si'(s)\right|_{\kappa_0=0}}{s-M^2}\,, \eeq which
is a special case of a `sideways dispersion relation' as is
demonstrated below.

The sideways dispersion for the \amm\ is obtained by
considering the half-off-shell electromagnetic vertex:
\bea
\eqlab{hovertex}
\Gamma^\mu (p',p)
&=& \La^{(+)}(\slap\,')\left[ F^{(+)}(s',q^2)\,\ga^\mu + G^{(+)}(s',q^2)\,\frac{(p+p')^\mu}{2M}\right]
\nn\\
&+&\La^{(-)}(\slap\,')\left[ F^{(-)}(s',q^2)\,\ga^\mu + G^{(-)}(s',q^2)\,\frac{(p+p')^\mu}{2M}\right]\,,
\eea
with $q=p'-p$ the photon momentum,  $p'$ the off-shell particle momentum ($s'={p'}^2$), and
$p$ the on-shell momentum ($p^2=M^2$). Furthermore,
\beq
\eqlab{proj}
\La^{(\pm)}(\slap)  = \frac{\slap \pm M}{2M}\,,
\eeq
are
the positive- and negative-energy state projections, and $F^{(\pm)}$ and $G^{(\pm)}$ are corresponding
half-off-shell form factors. In this decomposition of the vertex, the \amm\ is identified as
$\kappa= -G^{(+)}(M^2,0)$.

Akin to the Compton amplitudes considered above, the form factors
must be analytic functions of $s'$ everywhere in the complex plane
except along the unitarity cut, which lies on the real axis for
Re$\,s'>M^2$.  This analyticity allows a dispersive representation
for each of the form factors: \bea F^{(\pm)}(s',q^2) &= &
\frac{1}{\pi} \int\limits_{M^2}^\infty ds \,\frac{{\rm Im} \,
F^{(\pm)}(s,q^2)}{s-s'-i\eps} \,,\\
G^{(\pm)}(s',q^2) &= & \frac{1}{\pi} \int\limits_{M^2}^\infty
ds\,\frac{{\rm Im} \, G^{(\pm)}(s,q^2)}{s-s'-i\eps} \,, \eea and
these relations are called {\it sideways} dispersion
relations~\cite{bincer60,drell65}.

In particular, from the relation for $G^{(+)}$ we find \beq \kappa
= - \frac{1}{\pi} \int\limits_{M^2}^\infty ds\,\frac{{\rm Im} \,
G^{(+)}(s,0)}{s-M^2} \,, \eeq which is of the same form as the
linearized GDH sum rule, \Eqref{sranddr}.  Specifically, one can
show that \beq {\rm Im} \, G^{(+)}(s,0) =-\frac{M^2}{e^2}\, \left.
\De\si'(s)\right|_{\kappa_0=0} + \frac{M^2}{e^2}\, \left[
\De\si'(s)- \De\si_s'(s)\right]_{\kappa_0=0}\,, \eeq where the
second term integrates to 0. The function $\De\si_s'$ is defined
as the interference of the tree-level amplitude, \Figref{CStree},
and the $s$-channel graphs with a Pauli-coupling insertion (first
two graphs in \Figref{CStree0}). Recall that the function
$\De\si'$ is the interference of the tree-level graphs and  both
the $s$- and $u$-channel graphs with a Pauli-coupling insertion
(the four graphs in \Figref{CStree0}), and analogous relations for
the nucleon case will be discussed in the next section.

\section{Sum rules in chiral perturbation theory}

Consider now a theory of nucleons interacting with pions via
pseudovector coupling: \beq {\cal L}_{\pi NN} = \frac{g}{2 M}\,
\bar\psi \,\ga^\mu \,\ga^5 \,\tau^a \,\psi \,\pa_\mu \pi^a, \eeq
where $g$ is the pion-nucleon coupling constant, $M$ is the
nucleon mass, $\tau^a$ are isospin Pauli matrices, $\psi$ is the
nucleon field and $\pi^a$ is the isovector pion field. For our
purposes this Lagrangian is sufficient to obtain the leading order
results of chiral perturbation theory.

To lowest order in the coupling $g$, the photoabsorption cross
section in this theory is dominated by the single pion
photoproduction graphs as displayed in \Figref{Born_chpt}(a), and
we find the corresponding helicity difference cross sections~:
\begin{subequations}
\eqlab{piprod}
\bea
\De\si^{(\pi^0 p)}(\w) &=&
\frac{\pi C}{M^2 x^2}\,\left\{ (2\al \bar s+1-x) \ln\frac{\al+\la}{\al-\la} -
2 \la[ x (\al-2) + \bar s (\al+2) ] \right\}, \qquad \\
\De\si^{(\pi^+ n)} (\w)&=& \frac{2\pi C}{M^2 x^2}\,
\left\{  -\mu^2 \ln\frac{\be+\la}{\be-\la} + 2\la(\bar s\be-x\al)\right\}, \\
\De\si^{(\pi^0 n)} (\w)&=& 0,\\
\De\si^{(\pi^- p)} (\w)&=&
\frac{2\pi C}{M^2 x^2}\,\left\{ -\mu^2 \ln\frac{\be+\la}{\be-\la} + (2\al \bar
s-1-x) \ln\frac{\al+\la}{\al-\la} -
2\bar s\la\right\},
\eea
\end{subequations}
which are expressed here in terms of the dimensionless quantities:
\begin{subequations}
\eqlab{notation}
\bea
&& C=\left(eg/4\pi\right)^2, \\
&& x=\w/M,\,\,\,\,\, \mu = m_\pi/M,\,\,\,\,\,\bar s = s/M^2 =1 +2x,\\
&& \al=(s+M^2-m_\pi^2)/2s,\\
&& \be =(s-M^2+m_\pi^2)/2s=1-\al ,\\
&& \la = (1/2s)\sqrt{s-(M+m_\pi)^2}\sqrt{s-(M-m_\pi)^2} \,.
\eea
with $m_\pi$ denoting the pion mass.
\end{subequations}

The nucleon anomalous magnetic moment is generated from loop
diagrams and hence begins at ${\cal O}(g^2)$, implying that the
$lhs$ of the GDH SR begins at ${\cal O}(g^4)$. Since the
tree-level cross sections above are ${\cal O}(g^2)$, we must
require, as in the case of QED, the consistency conditions
 \beq
\eqlab{conc1} \int\limits_{\w_{\rm th}}^\infty {d\omega\over
\omega }\, \De\sigma^{(I)} (\w) =0, \,\,\,\, \mbox{for $I=\pi^0
p,\,\pi^+ n,\, \pi^0 n, \pi^- p$}, \ \eeq where $\w_{\rm th} =
m_\pi (1+ m_\pi/2M)$ is the threshold of the pion photoproduction
reaction. This requirement is indeed verified for the
expressions given in \Eqref{piprod}---the consistency of GDH SR is
maintained in this theory for {\it each} of the pion production
channels.

We now turn our attention to the linearized GDH sum
rule---\Eqref{linsr}. In this case we introduce Pauli moments
$\kappa_{0p}$ and $\kappa_{0n}$ for both the proton and neutron,
respectively. The dependence of the resulting cross-sections on
these quantities can then generally be presented as: \bea
\De\si(\w;\, \kappa_{0p},\, \kappa_{0n}) &=&\De\si(\w)
+\kappa_{0p} \,\De\si_{1p}(\w) + \kappa_{0n}\, \De\si_{1n}(\w) +
\ldots\,, \eea where we denote \bea
 \De\si_{1i} &=& \left. \frac{\pa}{\pa \kappa_{0i}} \De\si(\w;\, \kappa_{0p},\, \kappa_{0n})
\right|_{\kappa_{0p}=\kappa_{0n}=0}\,. \eea Furthermore, we
introduce total proton and neutron photoproduction cross sections
$\De\si^{(p)}$ and $\De\si^{(n)}$ and write the corresponding GDH
SRs and their first derivatives.  In this way we obtain the
relations:
\begin{itemize}
\item[(i)]  the GDH SRs:
\begin{subequations}
\eqlab{newsrs}
\beq
\frac{2 \pi \al_{em}}{M^2}\,\kappa_p^2 =\frac{1}{\pi}\int\limits_{\w_{th}}^\infty
\!
\frac{d\w}{\w}\,\De\si^{(p)},\hspace{1.5cm}
\frac{2 \pi \al_{em}}{M^2}\,\kappa_n^2 =\frac{1}{\pi}\int\limits_{\w_{th}}^\infty
\!
\frac{d\w}{\w}\,\De\si^{(n)},
\eeq
\item[(ii)]  the {\it linearized} SRs
(valid to leading order in the coupling $g$):
\bea
\eqlab{linearsrp}
\frac{4 \pi \al_{em}}{M^2}\, \de \kappa_p  &=&
\frac{1}{\pi}\int\limits_{\w_{th}}^\infty \!
\frac{d\w}{\w}\,\De\si_{1p}^{(p)} , \\
\frac{4 \pi \al_{em}}{M^2}\, \de \kappa_n &=&
\frac{1}{\pi}\int\limits_{\w_{th}}^\infty \!
\frac{d\w}{\w}\,\De\si_{1n}^{(n)}.
\eea
 \item[(iii)] the consistency conditions (valid to leading order in the
coupling $g$): \beq \eqlab{ncons} 0 =
\frac{1}{\pi}\int\limits_{\w_{th}}^\infty \!
\frac{d\w}{\w}\,\De\si_{1n}^{(p)} ,\hspace{1.5cm}
0  = \frac{1}{\pi}\int\limits_{\w_{th}}^\infty \!
\frac{d\w}{\w}\,\De\si_{1p}^{(n)}. \eeq

\end{subequations}
\end{itemize}

The first derivatives of the cross-sections that enter in
\Eqref{newsrs}, to leading order in $g$, arise through the
interference of Born graphs \Figref{Born_chpt}(a) with the graphs
in \Figref{Born_chpt}(b) and we find:
\begin{subequations}
\eqlab{piprod1}
\bea
\De\si_{1p}^{(p)}\equiv \De\si_{1p}^{(\pi^0 p)}+ \De\si_{1p}^{(\pi^+ n)}&=&
\frac{\pi C}{M^2x^2}\left\{ 2x\la [4+(1 -2\al) (1+2\bar s)]
-2 \bar s\la (\al+2) \right.\nn\\
&-& \left. \mu^2x \ln\frac{\be+\la}{\be-\la}+(2\al \bar s+1-x)
 \ln\frac{\al+\la}{\al-\la} \right\},\\
\De\si_{1n}^{(n)}\equiv\De\si_{1n}^{(\pi^0 n)}+ \De\si_{1n}^{(\pi^- p)}&=&
\frac{\pi C}{M^2x}\left\{ 2\la
+\mu^2\ln\frac{\be+\la}{\be-\la} -\ln\frac{\al+\la}{\al-\la}
\right\}, \qquad\\
\De\si_{1n}^{(p)}\equiv \De\si_{1n}^{(\pi^0 p)}+ \De\si_{1n}^{(\pi^+ n)}&=&
\frac{2\pi C}{M^2x^2}\left\{\ln\frac{\al+\la}{\al-\la} +2\la (x\be -\bar s \al)
\right\},\\
\De\si_{1p}^{(n)}\equiv\De\si_{1p}^{(\pi^0 n)}+ \De\si_{1p}^{(\pi^- p)}&=&
\frac{2\pi C}{M^2x^2}\left\{(2\bar s\al-x) \, \ln\frac{\al+\la}{\al-\la} +2\la
(x -2\bar s ) \right\}.
\eea
\end{subequations}
Using the latter two expressions we easily verify the consistency
conditions given in \Eqref{ncons}.

Employing the linearized
SRs, we obtain for the terms independent of $\kappa_{0p}$ and $\kappa_{0n}$:
\begin{subequations}
\eqlab{amms} \bea \de\kappa_p &=& \frac{M^2}{\pi e^2}\int\limits_{\w_{\mathrm{th}}}^{\infty}\!
\frac{d\w}{\w} \De\si_{1p}^{(p)}\nn\\
&=&\frac{g^2}{(4\pi)^2 } \left\{1 -
  \frac{\mu \,\left( 4 - 11{\mu }^2 + 3{\mu }^4 \right) }{\sqrt{1 - \quarter
{\mu }^2}}
  \arccos \frac{\mu }{2} - 6{\mu }^2+
  2{\mu }^2\left( -5 + 3\,{\mu }^2 \right) \ln \mu \right\}, \\
\de\kappa_n &=&\frac{M^2}{\pi e^2}
\int\limits_{\w_{\mathrm{th}}}^{\infty}\! \frac{d\w}{\w}
\De\si_{1n}^{( n)} =\frac{-2g^2}{(4\pi)^2 } \left\{2 -
\frac{\mu\,(2-\mu^2)}{\sqrt{1 - \quarter {\mu }^2}} \arccos
\frac{\mu }{2} - 2\mu^2 \ln \mu \right\}. \label{eq:bh}\eea
\end{subequations}
We have checked that the expressions of \Eqref{amms} agree with the one-loop
calculation done by using the standard Feynman-parameter
technique. To this order, the {\it pseudoscalar} pion-nucleon
coupling gives exactly the same result, which can easily be
verified by using the expressions of Appendix B.

We note that the chiral expansion of the corrections~\Eqref{amms} begins
with a constant, i.e., at ${\cal O}(1)$, and not with $m_\pi$  as is inferred by
power counting. However this is not a problem, because the power-counting-violating constant term
is absorbed into the counter term $\kappa_0$. The rest of the loop contribution obeys power counting.
In general, in relativistic ChPT the power counting applies only to
{\it renormalized} loop contributions, where the naively power-counting violating terms are absorbed 
into counter terms~\cite{Geg99}.

Even after such a renormalization of the constant terms in~\Eqref{amms},
 our sum-rule result  is {\it not} in agreement with the
covariant ChPT calculation of Kubis and Mei\ss ner~\cite{Kubis:2000zd}, which is based upon the {\it infrared
regularization} procedure of Becher and Leutwyler. The differences appear
only in the terms that are analytic in the quark mass, $m_q \sim m_\pi^2$.
This discrepancy can be traced back to 
 the fact that the 'infrared-regularized' loop
amplitudes do not satisfy the usual dispersion relations --- their
analytic properties in the energy plane are complicated by an
additional cut due to an explicit dependence on $\sqrt{s}$. In
other words, they do not obey the analyticity constraint
which is imposed on the sum rule calculation. 
 Certainly,
as the differences are analytic in the quark mass, they can be reconciled
order by order due to the inclusion of the appropriate counter terms.

\begin{figure}[h,b,t,p]
\centerline{
\epsfxsize=13cm
\epsffile{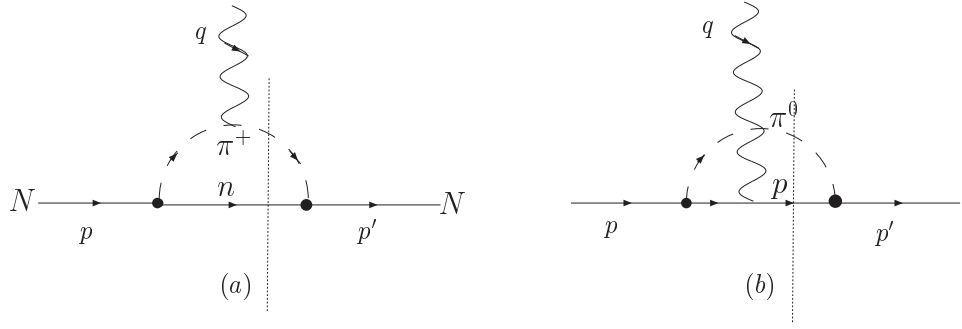}
}
\caption{Absorptive part of the $\gamma N N$ vertex
with final nucleon off the mass shell, used in the  sideways dispersion relation for the nucleon \amm.
Diagram (a) : $\pi^+ n$ loop where the photon couples to the $\pi^+$;
diagram (b) : $\pi^0 p$ loop where the photon couples to the charge of the
proton.
The vertical dotted lines indicate that the $\pi N$
intermediate state is taken on shell.}
\label{fig:gannabs}
\end{figure}

The interpretation of the linearized GDH sum rule is clarified by
observing that it has the  form of a sideways dispersion relation.
As remarked above for the QED case,  one can write down the
following sideways dispersion relation for the loop contribution
to the nucleon \amm: \beq \eqlab{sidegp} \de \kappa_N =  -
\frac{1}{\pi} \!\! \int\limits_{(M+m_\pi)^2}^\infty
\!\!\!ds\,\frac{{\rm Im} \, G_N^{(+)}(s,0)}{s-M^2} \,, \eeq where
$G_N^{(+)}$ is now the appropriate half-off-shell form factor for
the nucleon, cf.~\Eqref{hovertex}. The absorptive part of this
form factor to one loop is obtained by computing the indicated
cuts of the graphs in \Figref{gannabs}. For instance, the result
for the proton can be written as:
\begin{eqnarray}
\mathrm{Im} \, G_p^{(+)} = -\frac{M^2}{e^2}\left[
\De\si_{s1p}^{(\pi^+n)} + \De\si_{s1p}^{(\pi^0 p)} \right] ,
\end{eqnarray}
where the first term is equivalent to the contribution of
\Figref{gannabs} (a), whereas the second term represents
\Figref{gannabs} (b).  Explicit forms of these functions
$\De\si_s$ are listed in the Appendix B.  Making the connection
with the linearized GDH sum rule, we find that these $\De\si_s$
can equivalently be determined by computing the interference of
the Born pion-photoproduction graphs (\Figref{Born_chpt}) and  the
$s$-channel graph with a Pauli-coupling insertion [first graph in
\Figref{Born_chpt}(b)].

Note that $\De\si_{s1p}^{(\pi^+ n)}$ entering the above sideways
dispersion relation is exactly the same as the corresponding
integrand $\De\si_{1p}^{(\pi^+ n)}$ entering the linearized GDH
sum rule of Eq.~(\ref{eq:linearsrp}). The term
$\De\si_{s1p}^{(\pi^0 p)}$ in the integrand of
Eq.~(\ref{eq:sidegp}) on the other hand is different from the
corresponding term $\De\si_{1p}^{(\pi^0 p)}$ in
Eq.~(\ref{eq:linearsrp}). However, one can easily check that upon
integration both integrands give the same contribution to the
magnetic moments.  We have verified this result for both  pseudo-scalar  and pseudo-vector $\pi NN$
couplings.

In Appendix C, we, for completeness, list all the expressions for the {\it second}
derivatives of the GDH cross-sections for different single-pion production channels in 
Born approximation.

\section{Chiral corrections to nucleon polarizabilities}
It is well-known that the Heavy-Baryon ChPT (HBChPT) at order
${\cal O}(p^3)$ yields the prediction for the electric and
magnetic polarizabilities of the nucleon:
\begin{subequations}
\begin{eqnarray}
\eqlab{alphabet}
&&\alpha_E^{(HBLO)}=\frac{5 \pi \al_{em} }{6\, m_\pi} \left( \frac{g_A}{4
\pi f_\pi}\right)^2
=12.2 \times 10^{-4}~\mbox{fm}^3,\\
&&\beta_M^{(HBLO)}=\frac{\pi \al_{em}}{12\, m_\pi}
\left( \frac{g_A}{4 \pi f_\pi}\right)^2 =1.2 \times 10^{-4}~\mbox{fm}^3\,,
\end{eqnarray}\label{mei}
\end{subequations}
where $g_A\simeq 1.26$, $f_\pi \simeq 92.4$ MeV.  Here the
couplings are related to the $\pi NN$ coupling constant used in
the previous section via the Goldberger-Treiman relation:
$g_A/f_\pi = g/M$.

Eqs. \ref{mei} are a true prediction of ${\cal O }(p^3)$ HBChPT
(there are no counter-terms at this order) and turn out to be in
remarkable agreement with experiment, {\em e.g.}, for the proton
\cite{polz}:
\begin{subequations}
\begin{eqnarray}
\alpha_E^{(exp)}&=&12.1\pm 0.3 \,\mbox{(stat)}\mp 0.4 \,\mbox{(syst)} \pm 0.3\, \mbox{(mod)}
\,\,[\times 10^{-4}\,\,{\rm
fm}^3]\, ,\\
\beta_M^{(exp)}&=& 1.6 \pm 0.4 \,\mbox{(stat)}\pm  0.4 \,\mbox{(syst)} \pm 0.4\, \mbox{(mod)}
\,\,[\times 10^{-4}\,\,{\rm
fm}^3] \,.
\end{eqnarray}
\end{subequations}

By using forward sum rules, we can discuss here only the {\it sum}
of the electric and magnetic polarizabilities.  On the
experimental side, a recent determination from the Baldin's sum
gives~\cite{Bab98}:
\begin{eqnarray}
 \label{DDeq2.2.18}
(\alpha_E+\beta_M)_p^{exp} & = &
(13.69\pm0.14)\times 10^{-4}~{\rm{fm}}^3\ , \nonumber \\
(\alpha_E+\beta_M)_n^{exp} & = & (14.40\pm0.66)\times
10^{-4}~{\rm{fm}}^3\ ,
\end{eqnarray}
for proton and neutron, respectively.

In order to find the leading order relativistic prediction of
chiral loops we have computed the unpolarized total
cross-sections, corresponding to the Born graphs of single-pion
photoproduction~:
\begin{subequations}
\bea
\si^{(\pi^0 p)} &=&
\frac{\pi C}{M\w^3}\,\left\{ [\w^2-\mu^2 \al\, s]\,
\ln\frac{\al+\la}{\al-\la} + 2 \la \left[ \w^2 (\al-2) 1 + s \mu^2  \right]
\right\}, \nn\\
\si^{(\pi^+ n)} &=& \frac{2\pi C}{M\w^3}\,
\left\{ -\be \, s\,\mu^2\ln\frac{\be+\la}{\be-\la} +2\la
\, (\al\,\w^2 +s \,\mu^2)\right\}, \\
\si^{(\pi^0 n)} &=& 0,\nn\\
\si^{(\pi^- p)} &=&
\frac{2\pi C}{M\w^3}\,\left\{
\w^2 \ln\frac{\al+\la}{\al-\la}-\mu^2(s\be-\mu^2M^2)
\,\ln\frac{\be+\la}{\be-\la}\, \frac{\al+\la}{\al-\la} + 2 s\mu^2\la \right\}.
\eea
\end{subequations}
Substituting these expressions into the Baldin SR,
Eq.~(\ref{eq:baldinsr}), we obtain:
\begin{subequations}
\eqlab{alberesult}
\bea
(\al_E+\be_M)_p^{(RLO)}&=&\frac{e^2g^2}{16\pi^3 M^3}\left\{[3(1-4\mu^2+2\mu^4)
+\mbox{$\frac{1}{3}$}\mu^2]\ln\mu+\frac{406-737\mu^2+304\mu^4-36\mu^6}{6(4-\mu^2
)^2}
\right.\nn\\
&+&\left.
\frac{44-788\mu^2+1500\mu^4-899\mu^6+215\mu^8-18\mu^{10}}{3\mu(4-\mu^2)^{5/2}}\,
\arctg{\sqrt{\frac{4}{\mu^2}-1} } \,\right\}\,,\\
(\al_E+\be_M)_n^{(RLO)}&=&\!\frac{e^2g^2}{16\pi^3 M^3}\left\{
\ln\mu \right. \nn \\
&+& \left.
\frac{1}{(4-\mu^2)^2}\left[\frac{2(2-3\mu^2)(11-5\mu^2)-3\mu^6}{3\mu\sqrt{4-\mu^
2}}\,
\arctg{\sqrt{\frac{4}{\mu^2}-1} } +5-\mu^2\right]\right\}.\nn\\
\eea
\end{subequations}
Note that an identical result is obtained in the conventional
one-loop Feynman diagram calculation~\cite{BKMalbeloop,Metz96}.
The corresponding (chiral) $\mu$-expansion reads:
\begin{subequations}
\eqlab{albeexpand}
\bea
(\al_E+\be_M)_p^{(RLO)}&=&\frac{e^2g^2}{(4\pi)^2 M^3}
\frac{11}{48\mu}\left\{ 1+\frac{48 (4+3\ln\mu)}{11\pi}\mu - \frac{1521}{88}\mu^2
 +\ldots\right\}, \\
(\al_E+\be_M)_n^{(RLO)} &=&\frac{e^2g^2}{(4\pi)^2 M^3}
\frac{11}{48\mu}\left\{ 1+\frac{4(1+12\ln\mu)}{11\pi}\mu - \frac{117}{88}\mu^2 +
\ldots\right\},
\eea
or, numerically (using $g^2/4\pi=13.8$, $M=0.9383$ GeV, $\mu=0.148$),
\bea
(\al_E+\be_M)_p^{(RLO)}&=& 14.5 -5.2 - 5.5 +\ldots= 3.8, \\
(\al_E+\be_M)_n^{(RLO)}&=& 14.5 - 5.4 - 0.4 + \ldots=8.7, \eea
\end{subequations}
in units of $10^{-4}\,\,{\rm fm}^3$.
As one can clearly see, the fully relativistic leading order
result is substantially different from the non-relativistic
(heavy-baryon) limit. The higher order corrections, which are
suppressed by $m_\pi/M\simeq 1/7$, and hence are expected to be
small, appear with large coefficients and generate a substantial
modification of the leading order result. This (relativistic)
effect is likely to allow one to accommodate the relatively large
$\De$-resonance contribution to the magnetic
polarizability~\cite{Pas05}.

For the forward spin polarizability, Bernard, Hemmert and Mei\ss ner obtain\cite{bhm}~: \beq \gamma_0^{(p)} =\frac{e^2 g^2}{16\pi^3
\mu^2 } \left[1 - \frac{21\,\pi \,\mu }{8} -
 \left( 20  + 26\,\ln\mu
     \right) {\mu }^2  + \frac{1875\,\pi \,{\mu }^3}{64} + \ldots
\right] ,\eeq while, using the sum rule of Eq.~\ref{eq:gamma0sr},
we obtain~: \beq 
\eqlab{spinpolza}\gamma_0^{(p)}= \frac{e^2 g^2}{16\pi^3 \mu^2 }
\left[1 - \frac{21\,\pi \,\mu }{8} -
 \left( \frac{59}{2}  + 26\,\ln\mu
     \right) {\mu }^2 + \frac{1875\,\pi \,{\mu }^3}{64} + \ldots
\right]. \eeq The difference between both expressions is confined
to the analytic terms. The sum rule calculation has analyticity
built in explicitly.

\section{Chiral extrapolations}
It  is instructive to examine the chiral behavior of the one-loop
result for the nucleon magnetic moment. Expanding \Eqref{amms}
around the chiral limit ($m_\pi=0$) we have 
\begin{subequations}
 \eqlab{expand_amms}\bea 
\kappa_p^{\rm (loop)} &=& \frac{g^2}{(4\pi)^2 }\left\{1 -2\pi
\mu-2\,(2+5\ln\mu)\,\mu^2+\frac{21\pi}{4}\,\mu^3
+ O(\mu^4)\right\}, \\
\kappa_n^{\rm (loop)} &=&\frac{g^2}{(4\pi)^2 } \left\{-4 +2\pi
\mu-2\,(1-2\ln\mu)\,\mu^2-\frac{3\pi}{4}\,\mu^3+ O(\mu^4)
\right\}. \eea 
\end{subequations}
Apart from the first term, the constant, which renormalizes the counter term as described above, 
this expansion corresponds with the heavy-baryon expansion.
The term linear in pion mass (recall that
$\mu=m_\pi/M$) is the well-known leading nonanalytic (LNA)
correction.  

On the other hand, expanding the same expressions
around the large $m_\pi$ limit
we find
\begin{subequations}
\bea
\kappa_p^{\rm (loop)} &=& \frac{g^2}{(4\pi)^2 }  \,(5-4\ln\mu)\frac{1}{\mu^2}
+ O(\mu^{-4}), \\
\kappa_n^{\rm (loop)} &=&\frac{g^2}{(4\pi)^2 } \,2 (3-4\ln\mu)\frac{1}{\mu^2} +
O(\mu^{-4}).
\eea
\end{subequations}
What is intriguing here is that the one-loop
correction to the nucleon \amm\ for heavy quarks  behaves as
$1/m_{quark}$ (where $m_{quark} \sim m_\pi^2)$, precisely as expected from a
constituent quark-model picture~\cite{Leinweber:1998ej}.  Here this is a result of subtle cancellations
in \Eqref{amms} taking place for large values of $m_\pi$. In contrast, the
infrared regularization procedure~\cite{Kubis:2000zd} gives
the result which exhibits pathological
behavior with increasing pion mass and diverges for $m_\pi =2M$.

Since the expressions in \Eqref{amms} both have the correct large
{\it and} small $m_\pi$ behavior they should be better suited for
the chiral extrapolations of lattice results than the usual
heavy-baryon expansions or the ``infrared-regularized''
relativistic theory. This point is clearly demonstrated by
\Figref{chibehavior}, where we plot the $m_\pi$-dependence of the
full [\Eqref{amms}], heavy-baryon, and
infrared-regularization\cite{Kubis:2000zd} leading order result
for the magnetic moment of the proton and the neutron, in
comparison  to recent lattice data\cite{Zan04}.  In presenting
these results we have added a constant shift (counter-terms
$\kappa_0$) to the magnetic moments, i.e., \bea
\mu_p&=&(1+\kappa_{0p}+\kappa_p^{\rm (loop)})(e/2M),\\
\mu_n&=&(\kappa_{0n}+\kappa_n^{\rm (loop)}) (e/2M) \eea and fitted
them to the known experimental value of the magnetic moments at
the physical pion mass, $\mu_p\simeq 2.793$ and $\mu_n\simeq
-1.913$, shown by the open diamonds in the figure. For the value
of the $\pi NN$ coupling constant we have used $g^2/4\pi = 13.5$.
The $m_\pi$-dependence away from the physical point is then a
prediction of the theory. The figure clearly shows that the
SR-motivated extrapolation, shown by the dotted lines, is in
a good agreement with the $m_\pi$-dependence obtained in lattice
gauge simulations.

 It is therefore tempting to use the SR results for the parametrization
of lattice data. For example, we consider the following elementary
two-parameter form:
\begin{subequations}
\eqlab{paran}
\bea
\mu_p&=&\left(1+\frac{\tilde \kappa_{0p}}{1+a_p m_\pi^2}+\kappa_p^{\rm (loop)}
\right)\frac{e}{2M},\\
\mu_n&=&\left(\frac{\tilde\kappa_{0n}}{1+a_n
m_\pi^2}+\kappa_n^{\rm (loop)}\right)\frac{e}{2M}\,, \eea
\end{subequations}
where $\tilde \kappa_{0p}$ and $\tilde \kappa_{0n}$ are fixed to
reproduce the experimental magnetic moments at the physical
$m_\pi$. The parameter $a$ can be fitted to lattice data. The
solid curves in \Figref{chibehavior} represent the result of such
a single parameter fit to the lattice data of Ref.~\cite{Zan04}
for the proton and neutron respectively, where $a_p=1.6/M^2$ and
$a_n=1.05/M^2$, and $M$ is the physical nucleon mass.  
This parametrization incorporates the experimental value at the physical
pion mass and reproduces very well the 
trend of the lattice data.

We would like to stress here that at larger pion masses the ChPT result should be interpreted as {\it merely 
a convenient parametrization of lattice data, not as a prediction of the theory.}
In this way one obtains a parametrization which fits the lattice data at larger values of $m_\pi$ and has the correct
dependence at lower values of $m_\pi$. Apparently, the relativistic ChPT result which is
consistent with analyticity provides the most convenient ground for the fit to lattice data, because it has an $1/m_q$ behavior at larger values of $m_\pi$.
It is useful to test the consistency of lattice simulations at larger $m_\pi$ and the
experimental values based on such a single-free-parameter form which encodes the correct
chiral behavior at low pion masses.

The importance of the relativistic effects is even more emphasized in the polarizabilities.
In \Figref{polzachib} 
we have plotted the ratio of one-loop $\alpha_E +\beta_M$, \Eqref{alberesult}, 
to  its leading non-analytic term [first term in the expansion \Eqref{albeexpand}]. 
Recently, the electric and magnetic polarizabilities of hadrons have been 
calculated in lattice QCD 
using the external field method~\cite{Christensen:2004ca,Zhou:2002km}. 
This method amounts to extract the electric and magnetic 
polarizabilities of hadrons from the quadratic term in the mass shift 
in an external electromagnetic field.  
The low statistics, pioneering results have so far been obtained for 
pion masses above 450 MeV. As higher statistics calculations will become 
available in the near future~\cite{Lee05, Zanotti05},  
our sum rule calculation 
of the pion mass dependence of the forward polarizabilities 
provides a way to connect these next generation lattice results to experiment.

\section{Prospects for the future}

In this section we outline a procedure for two-loop calculations
of the \amm\ by using derivative forms of the GDH SR.  Again, the
GDH SR in its original form is practically unsuitable for such
calculations, because in order to compute $\kappa$ in, {\em e.g.},
QED to order $\al_{em}^2$ one needs to compute the photoabsorption
cross-sections to order  $\al_{em}^5$---also to two loops!  By
contrast, using the derivative trick of Sect.~III, we will need
the (derivatives of) cross-sections up to order $\al_{em}^3$
only---one loop forms.

Introducing the `trial' value of $\kappa_0$, such that
the full \amm\ is defined as $\kappa = \kappa_0 + \de \kappa$, and the GDH cross-section
can be written as
\beq
\De\si(\kappa_0) = \De \si + \kappa_0 \, \De\si' + \half \kappa_0^2 \, \De\si'' + \ldots\,,
\eeq
the first and second derivatives of the GDH SR at $\kappa_0=0$ read, respectively:
\begin{subequations}
\bea
\eqlab{first}
&& \de \kappa + \de \kappa\, \de\kappa' = \frac{M^2}{4\pi^2\al_{em}}
\int\limits_{0}^\infty \!\frac{d\w}{\w}\, \De\si' \\
\eqlab{second}
&& 2 \de \kappa' +(\delta \kappa')^2 +\de \kappa\, \de\kappa'' =\frac{M^2}{4\pi^2\al_{em}}
\int\limits_{0}^\infty \!\frac{d\w}{\w}\, \De\si''
\eea
\end{subequations}
Now, consider \Eqref{first} at order $\al_{em}^2$:
\beq
\eqlab{twoloopsr1}
\de \kappa^{(2)} + \de \kappa^{(1)}\, \de\kappa'^{(1)} = \frac{M^2}{4\pi^2\al_{em}}
\int\limits_{0}^\infty \!\frac{d\w}{\w}\, \De\si'^{(3)}\,,
\eeq
where the superscript indicates the order of $\al_{em}$ to which the quantity is considered.
We can use \Eqref{first} and \Eqref{second} at
order $\al_{em}$ to obtain, respectively:
\begin{subequations}\bea
\de \kappa^{(1)} &=& \frac{M^2}{4\pi^2\al_{em}}
\int\limits_{0}^\infty \!\frac{d\w}{\w}\, \De\si'^{(2)}\,, \\
\de \kappa'^{(1)} &=& \frac{M^2}{8\pi^2\al_{em}}
\int\limits_{0}^\infty \!\frac{d\w}{\w}\, \De\si''^{(2)}\,.
\eea
\end{subequations}
Substituting these in \Eqref{twoloopsr1} we find the {\it
linearized} GDH sum rule at order $\al_{em}^2$: \beq
\eqlab{linsr2} \de\kappa^{(2)} = \frac{M^2}{4\pi^2\al_{em}} \left[
\int\limits_{0}^\infty \!\frac{d\w}{\w} \, \De\si'^{\,(3)} -
\frac{M^2}{8\pi^2\al_{em}}\left( \int\limits_{0}^\infty
\!\frac{d\w}{\w} \, \De\si'^{\,(2)}\right)\,\left(
\int\limits_{0}^\infty \!\frac{d\w}{\w} \,
\De\si''^{\,(2)}\right)\right]\, , \eeq and, using this sum rule
one can obtain all two-loop corrections to the \amm\ by computing
the first derivative of the GDH cross-section $ \De\si'$ at
one-loop and tree level, and a second derivative of the GDH
cross-section $\De\si''$ at tree level.

For the QED case we have thus far worked out the required
derivatives at tree level only. The first derivative is given in
\Eqref{gdh2}, while the second derivative is given by: \beq
\De\si^{''(2)}=\frac{2\pi{\al_{em}^2 }}{M^2x}\,\left[ - \frac{9 +
29\,x + 18\,x^2 - 10\,x^3}
         {{\left( 1 + 2\,x \right) }^2} + \left( 1 +
         \frac{9}{2\,x} \right) \,\log (1 + 2\,x) \right]\,,
\eeq with $x=\w/M$.  As already shown, the dispersion integral
over the first derivative yields the Schwinger correction, {\it
cf.} \Eqref{Schresult}. The integral over the second derivative is
divergent and to evaluate it we introduce an ultraviolet cutoff
$\La \gg M$: \beq \frac{1}{\pi}\int\limits_{0}^\La
\!\frac{d\w}{\w} \, \De\si''^{\,(2)} = \frac{\al_{em}^2}{M^2}
\left(5 \log \frac{2\La}{M} - 6\right)\,. \eeq The same cutoff
needs to be introduced in the first integral of the sum rule
\Eqref{linsr2}, such that the total result is independent of the
cutoff. Thus, in QED we obtain \beq \eqlab{QEDlinsr2}
\de\kappa^{(2)} = \frac{M^2}{4\pi^2\al_{em}} \int\limits_{0}^\La
\!\frac{d\w}{\w} \, \De\si'^{\,(3)} -
  \frac{\al_{em}^2}{16\pi^2} \left(5 \log \frac{2\La}{M} - 6\right)\,.
\eeq

In order to complete this two-loop calculation of the electron
\amm\ in QED one would need to evaluate the first derivative
$\De\si'$ at order $\al_{em}^3$ which is given by the interference
of the tree-level and one-loop Compton scattering amplitudes, as
well as by the tree-level bremsstrahlung and pair production
mechanisms, all  with one insertion of the Pauli vertex.
Diagrammatically this calculation is depicted in
\Figref{qed2loop}. It is a future challenge to perform such
calculation and to verify whether the linearized GDH SR reproduces
the two-loop result obtained by usual
techniques~\cite{twoloopQED}: \beq \de \kappa^{(2)}=
\frac{\al_{em}^2}{\pi^2}\left[ \frac{197}{144} +
\frac{\pi^2}{2}(\sixth - \log 2) + \frac{3}{4}\, \zeta(3)\right]
\simeq  -0.328479\, (\al_{em}^2/\pi^2)\,. \eeq This would be an
exact test of the GDH SR at the two-loop level.

In the case of the nucleon, sum rules analogous to
\Eqref{twoloopsr1} may provide a more efficient method to do a
two-loop calculation for the nucleon \amm or for the forward spin
polarizability, since much is already known about the one-loop
ChPT amplitudes of pion-photoproduction~\cite{BKMLee}, which are
required  in such a calculation.

\section{Conclusion}

Direct application of QCD to low energy hadronic physics is made
challenging by the feature that the quark/gluon degrees of freedom
in terms of which QCD is couched are confined within hadronic
systems. Nevertheless, the chiral EFT of QCD---chiral perturbation
theory---allows a predictive description of the low-energy
hadronic reactions.  Also, by focusing on the chiral structure of
the underlying Lagrangian such EFT methods permit one to make a
link to lattice QCD calculations, even in the case where the mass
of the underlying Goldstone bosons is considerably heavier than
found experimentally.  These are the two fronts which at present
make the chiral EFT indispensable in relating QCD to low-energy
observables.  On the other hand there remain significant problems.
Indeed, most such calculations involving nucleons are carried out
at low orders in so-called heavy baryon chiral perturbation
theory, which involves an expansion in powers of the inverse
nucleon mass.  On the other hand, we have demonstrated above by
simple examples involving the nucleon magnetic moment and
polarizabilities, that manifestly relativistic calculations, which
include chiral corrections to all orders in $\mu=m_\pi/M_N$, do a
better job than the ``heavy-baryon'' ones on both fronts in
capturing the full implications of the theory.

For this reason it is important to have at one's disposal methods
which allow such all orders evaluation of experimental quantities.
This can in principle be achieved using conventional relativistic
chiral loop calculations, but the price in terms of the number of
diagrams which must be evaluated is high.  For example, a one-loop
evaluation of the nucleon polarizability involves fifty-two such
diagrams for the proton and twenty-two for the neutron\cite{KMB}!
Above we have presented an alternative approach, involving the use
of dispersion relations and their related sum rules.  The
calculations done in this work were based on
real-Compton-scattering sum rules, such as GDH and Baldin sum
rules.  However, the results are identical to what one would
obtain in the usual loop calculations,  provided no manipulations,
such as {\it infrared regularization} which change the analytic
structure  are made.  As is shown in the works of Gegelia {\it et
al.}~\cite{Geg99}, there is no problem with power counting in this
straightforward formulation of covariant ChPT, if the
renormalization is performed in a suitable way. 
Both the infrared-regularized and the straightforward formulation
of relativistic baryon ChPT resum the nominally higher-order 
relativistic effects, albeit differently. We stress that
the resummation in the straightforward version 
is done in accordance with the principle of {\it analyticity}.

We demonstrated above the utility of taking derivatives of the GDH
sum rule, in order to convert it to forms which are sometimes
more calculationally robust.  We
showed that in some cases these modified versions of the sum rules
are equivalent to {\it sideways} dispersion relations, which
involve taking the final (or initial) nucleon four--momentum
off-shell.  Using such relations, we straightforwardly evaluated
fully relativistic nucleon magnetic moments and polarizabilities,
using only tree level inputs and showed that they were fully
equivalent to values obtained via much more labor-intensive loop
techniques.  We showed that such methods should also permit a
future evaluation of the two loop electron anomalous magnetic
moment.

The successes demonstrated above suggest an obvious direction for
future work.  In particular, the use of dispersive methods would
seem to be ideal for a simplified way to perform two-loop
evaluations.  An obvious first test would be to use one-loop
corrected photoproduction amplitudes in order to produce
corresponding one-loop photoabsorption cross sections, which can
in turn be used to generate two-loop values of polarizabilities
and of magnetic moments which are relativistically correct.  These
can then be compared to ongoing calculations at two loop in the
heavy baryon expansion for the forward spin polarizability and
with the ``experimental" value obtained from the sum rule.   In
conclusion there is much which can be done with such methods, and
many future applications can be envisioned.

\section*{Acknowledgements}
This work is supported in part by DOE grant no.\ DE-FG02-04ER41302
and contract DE-AC05-84ER-40150 under which the Southeastern
Universities Research Association (SURA) operates the Thomas
Jefferson National Accelerator Facility.  The work of BRH is also
supported in part by the National Science Foundation under award
PHY02-44801.

\appendix

\section{Calculation of photoproduction cross-sections}
We write the pion photoproduction amplitude in the following general form:
\beq
T_{\pi \ga} = \bar u (p') \, \sum_{i=1}^4 A_i(s,t)\, O_i(\veps) \, u(p)\,,
\eeq
where $A$'s are the invariant amplitudes and $O$'s are given by
\bea
O_1 &=& \ga_5\,,\nn\\
O_2 &=& \veps \cdot \ga\cdot q \,\ga_5\,,\nn\\
O_3 &=& \ga \cdot \veps \,\ga_5\,,\\
O_4 &=& \ga \cdot q \,\ga_5\,,\nn
\eea
with $\veps \cdot \ga\cdot q\equiv \half [\ga_\mu, \ga_\nu] \, \veps^\mu q^\nu$.

The total and double-polarized cross-sections are given respectively  by
\bea
\si & = & \frac{2\pi \la}{x} \sum_{i,j=1}^{4} \int^{1}_{-1} d(\cos \th) \, Y_{ij}\, A_i \, A_j, \\
\De\si & = & \frac{2\pi \la}{x} \sum_{i,j=1}^{4}  \int^{1}_{-1} d(\cos \th)\, Z_{ij}\, A_i \, A_j,
\eea
with $\th$ the scattering angle and
\bea
Y_{ij} &=& {\rm Tr} \left[ \La^{(+)}(\slap') \, O_i \,    \La^{(+)}(\slap)\, O_j\right] \,,\\
Z_{ij} &=& {\rm Tr} \left[ \La^{(+)}(\slap') \, O_i \,    \La^{(+)}(\slap)\,{\vec \Sigma}\cdot {\vec  p}\,
O_j\right]\,,
\eea
where $\vec \Sigma = \ga_0 \ga_5 \vec \ga $,
$\vec p$ is the 3-momentum of the incoming nucleon, and the projection
operator $\La^{(+)}$ is defined in \Eqref{proj}.
More explicitly, the traces are given by
\bea
Y &= & \left( \begin{array}{cccc}
\half \bar t & -x\xi & \xi & y-x \\
-x \xi & -2xy & x+y & 0\\
\xi & x+y & -2 +\half \bar t & -x\xi \\
y-x & 0 & -x \xi & -2 x y
  \end{array} \right), \\
Z &= & \left( \begin{array}{cccc}
0 & x\xi & -\xi & 0 \\
x \xi & 2xy & -x-y & 0\\
-\xi & -x-y & 2 -\half \bar t & x\xi \\
0 & 0 & x \xi & 0
  \end{array} \right) ,
\eea
where  $x=(s-M^2)/2M^2=\w/M$,
$y=(M^2-u)/2M^2$, $\xi = p'\cdot\veps=\la \sqrt{\bar s}\, \sin \th$, $\bar s =s/M^2$, $\bar t =t/M^2$
($s$, $t$ and $u$ are the
Mandelstam variables).

In the notations of \Eqref{notation} the expression for the total cross-section of pion photoproduction in
Born approximation, including the nucleon \amm\ terms, takes the following form:
\bea
\si^{(\pi^0 p)} &=&
\frac{\pi C}{M^2 x ^3}\,\left\{ [x^2(1+2\kappa_p +\half\kappa_p^2)-\mu^2 \al\, \bar s]\,
\ln\frac{\al+\la}{\al-\la} \right. \nn\\
&+& \left. 2 \la \left[ x^2 (\al-2) (1+\kappa_p)^2 + \bar s \mu^2 +\half \kappa_p^2\al\, \bar s \right] \right\} ,\nn\\
\si^{(\pi^+ n)} &=& \frac{2\pi C}{M^2 x^3}\,\left\{ -\be \, \bar s\,\mu^2\,\ln\frac{\
\be+\la}{\be-\la} +2\la
\, (\al\,x^2 +\bar s \,\mu^2) \right. \nn\\
&+& x^2\,\la\, \left[-2(\kappa_p+\kappa_n)\,\be
+(\kappa_p^2+\kappa_n^2)(x\be-\half\mu^2)\right]
\nn\\
&+&\left. x^2\, \kappa_p\, \kappa_n  \left[ \half \ln\frac{\al+\la}{\al-\la}
-\la(2-\al\,\bar s+2(1+x)\be-\mu^2)\right] \right\}, \\
\si^{(\pi^0 n)} &=& \frac{\pi C }{M^2 x}\kappa_n^2 \left\{ \half
\ln\frac{\al+\la}{\al-\la}
+[-4+(2+\bar s)\,\al\,]\,\la\right\},\nn\\
\si^{(\pi^- p)} &=&
\frac{2\pi C}{M^2 x^3}\,\left\{
x^2 \ln\frac{\al+\la}{\al-\la}-\mu^2(\bar s\,\be-\mu^2)
\,\ln\frac{\be+\la}{\be-\la}\, \frac{\al+\la}{\al-\la} + 2 \bar s\mu^2\la \right.\nn\\
&+&x^2  (\kappa_p+\kappa_n)\left(2\la-\ln\frac{\al+\la}{\al-\la}\right)
+x^2(\kappa_p^2+\kappa_n^2)(x\be-\half \mu^2)\nn\\
&+& \left. x^2 \kappa_p\kappa_n \left[(1+\be)\la +\frac{1}{2}\ln\frac{\al+\la}{
\al-\la}\right] \right\}.\nn
\eea

\section{First derivatives of GDH Born cross-sections of pion photoproduction}
In this appendix we present explicit expressions for the $\De \si_1$ cross-sections
for the  pion photoproduction on the proton.
\subsection{Pseudo-vector coupling}
Computing the Born graph contribution with the pseudo-vector $\pi NN$ coupling we obtain:
\bea
\De\si_{1p}^{(\pi^0 p)} &=&
\frac{\pi C}{M^2 x^2}\left\{ (2\al\bar s+1-x)
\ln\frac{\al+\la}{\al-\la} -2\la \,[x(\al-2-\mu^2)+\bar s (2+\al)] \right\} \nn\\
&=& \De\si^{(\pi^0 p)} +
\frac{2\pi C}{M^2 x} \, \la \, \mu^2 \\
\De\si_{1n}^{(\pi^0 p)}& =&\De\si_{2nn}^{(\pi^0 p)}=\De\si_{2pn}^{(\pi^0 p)}= 0\\
\De\si_{1p}^{(\pi^+ n)} &=&\frac{\pi C}{M^2 x}\,
\left\{  -\mu^2 \ln\frac{\be+\la}{\be-\la} + 2\la(\bar s\,\be-x\al)\right\}, \\
\De\si_{1n}^{(\pi^+ n)}&=&\De\si_{1p}^{(\pi^+ n)}-\frac{4\pi C}{M^2 x^2}\,\al\,\bar s\,\la \, .
\eea
where the notations are detailed in \Eqref{notation}, and $\De\si^{(\pi^0 p)}$ is given in
\Eqref{piprod}.

\subsection{Pseudo-scalar coupling}
In the case of pseudo-scalar $\pi NN$ coupling the expressions are different.
\bea
\De\si_{1p}^{(\pi^0 p)} &=&
\frac{\pi C}{M^2 x^2}\left\{ 2\la \,(\al-\bar s-3+\mu^2) +(3-\mu^2)\ln\frac{\al+\la}{\al-\la}\right\},  \\
\De\si_{1n}^{(\pi^0 p)}& =& \De\si_{2nn}^{(\pi^0 p)}= \De\si_{2pn}^{(\pi^0 p)} = 0\\
\De\si_{1p}^{(\pi^+ n)} &=&\De\si_{1n}^{(\pi^+ n)}=
\frac{4\pi C}{M^2 x} \la\, \be .
\eea
Note, however, the contribution of these cross-sections to the linearized GDH sum rule \Eqref{newsrs}
is exactly the same in
both the pseudo-vector and pseudo-scalar cases.

\subsection{Integrands of the sideways dispersion relation}

The integrands entering the sideways dispersion relations of \Eqref{sidegp}
for the proton \amm\ are  given by~:
\bea
\De\si_{s1p}^{(\pi^+ n)} &=&\De\si_{1p}^{(\pi^+ n)},\\
\De\si_{s1p}^{(\pi^0 p)}& = &
\frac{\pi C}{M^2 x} \left\{ 2\la\, (1+\be)-\ln\frac{\al+\la}{\al-\la}\right\}.
\eea

\section{Second derivatives of GDH Born cross-sections of pion photoproduction}
In this appendix we list the expressions for the second derivatives of the GDH cross section, defined as: 
\beq
 \De\si_{2ij} = \left. \frac{\pa}{\pa \kappa_{0i}}
\frac{\pa}{\pa \kappa_{0j}} \De\si(\w; \, \kappa_{0p},\, \kappa_{0n})\right|_{\kappa_{0p}=\kappa_{0n}=0}\,.
\eeq
Using the pseudo-vector coupling, we obtain:
\bea
\De\si_{2pp}^{(p)} &=& \De\si_{2nn}^{(n)} =
\frac{2\pi C}{M^2x}\left\{ \la (4- 2x\be- 2\al + \mu^2 )
 -\ln\frac{\al+\la}{\al-\la} \right\},\\
\De\si_{2nn}^{(p)}&=&\De\si_{2pp}^{(n)} =
\frac{2\pi C}{M^2x}\left\{ \la [2(\bar s+x)\al -2x +\mu^2]
- \ln\frac{\al+\la}{\al-\la}\right\}, \qquad\\
\De\si_{2pn}^{(p)}&=&\De\si_{2pn}^{(n)} =
\frac{\pi C}{M^2x}\left\{ 2\la [-2\be(1+x)+ \bar s\al-2 +\mu^2]
+ \ln\frac{\al+\la}{\al-\la}\right\}. \qquad
\eea
Using the pseudo-scalar coupling, we obtain:
\bea
\half \De\si_{2pp}^{(\pi^0 p)} &=& \De\si_{2pn}^{(\pi^+ n)} =
\frac{\pi C}{M^2 x} \left\{ 2\la\, (1+\be)-\ln\frac{\al+\la}{\al-\la}\right\}, \\
\De\si_{2pp}^{(\pi^+ n )} &=& 
\frac{2\pi C}{M^2 x} \la \,(\mu^2- 2x\be) = 
\frac{4\pi C}{M^2 x} \la \,(\be - \al x)\,, \\
\De\si_{2nn}^{(\pi^+ n )} &=&  \frac{2\pi C}{M^2 x} \left\{ \la\, (1+\be +\al\, \bar s)
-\ln\frac{\al+\la}{\al-\la}\right\} .
\eea

\newpage

 \begin{figure}[b,h,t,p, c]
\centerline{
  \epsfxsize=7cm
  \epsffile{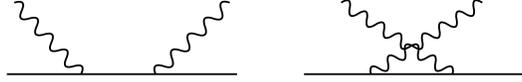}
}
\caption{Tree-level Compton scattering amplitude in QED.}
\figlab{CStree}
\end{figure}

 \begin{figure}[b,h,t,p, c]
\centerline{
  \epsfxsize=13cm
  \epsffile{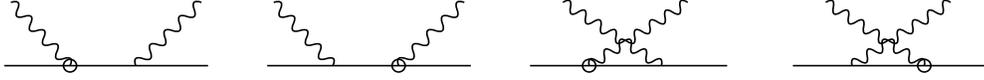}
}
\caption{Tree-level Compton scattering graphs with one Pauli-coupling insertion.
The circled vertex corresponds to the Pauli coupling.}
\figlab{CStree0}
\end{figure}

 \begin{figure}[b,h,t,p, c]
\centerline{
  \epsfxsize=13cm
  \epsffile{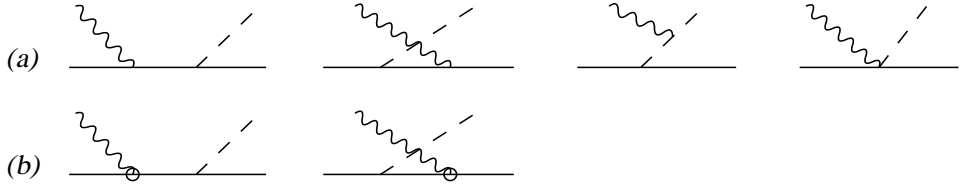}
}
\caption{Tree-level pion photoproduction graphs.
The circled vertex corresponds to the Pauli coupling.}
\figlab{Born_chpt}
\end{figure}

\begin{figure}[h,b,t,p]
\centerline{
  \epsfxsize=7cm
  \epsffile{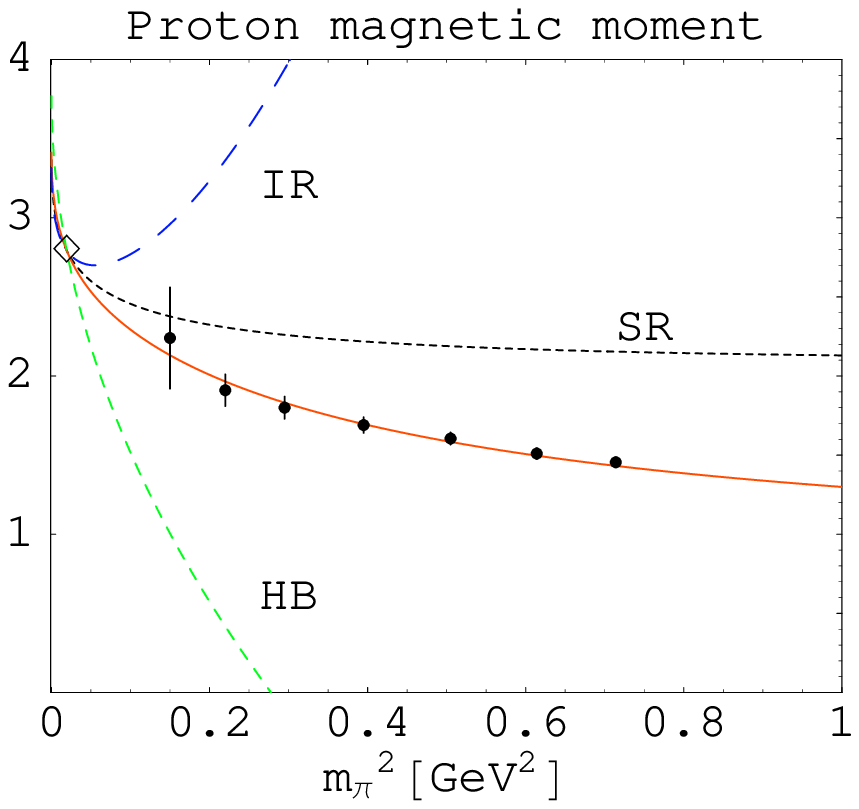}
\epsfxsize=7cm  \epsffile{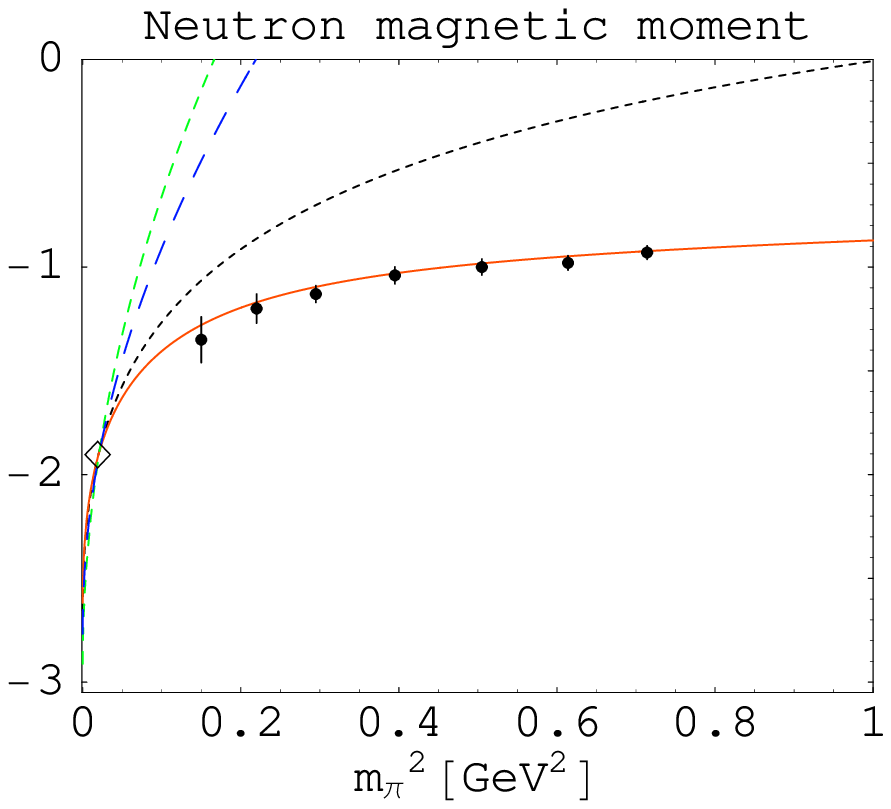}
}
\noindent \caption{ Chiral behavior of proton and neutron magnetic
moments (in nucleon magnetons) to one loop compared with lattice
data. ``SR'' (dotted lines): our one-loop relativistic result,
``IR'' (blue long-dashed lines): infrared-regularized relativistic
result, ``HB'' (green dashed lines): LNA term in the heavy-baryon
expansion. Red solid lines:  single-parameter  fit based on our SR
result. Data points are results of lattice simulations. The open
diamonds represent the experimental values at the physical pion
mass.} \figlab{chibehavior}
\end{figure}
\begin{figure}[h,b,t,p]
\centerline{
  \epsfxsize=8cm
  \epsffile{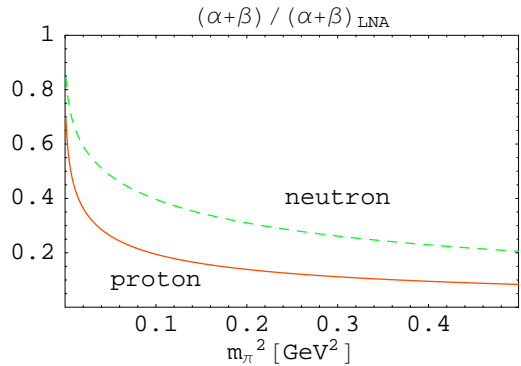}
}
\noindent \caption{ The ratio of the one-loop proton (solid curves) and neutron (dashed curves)
forward polarizabilities to their LNA terms in the heavy-baryon
expansion.} \figlab{polzachib}
\end{figure}

 \begin{figure}[b,h,t,p, c]
\centerline{
  \epsfxsize=15cm
  \epsffile{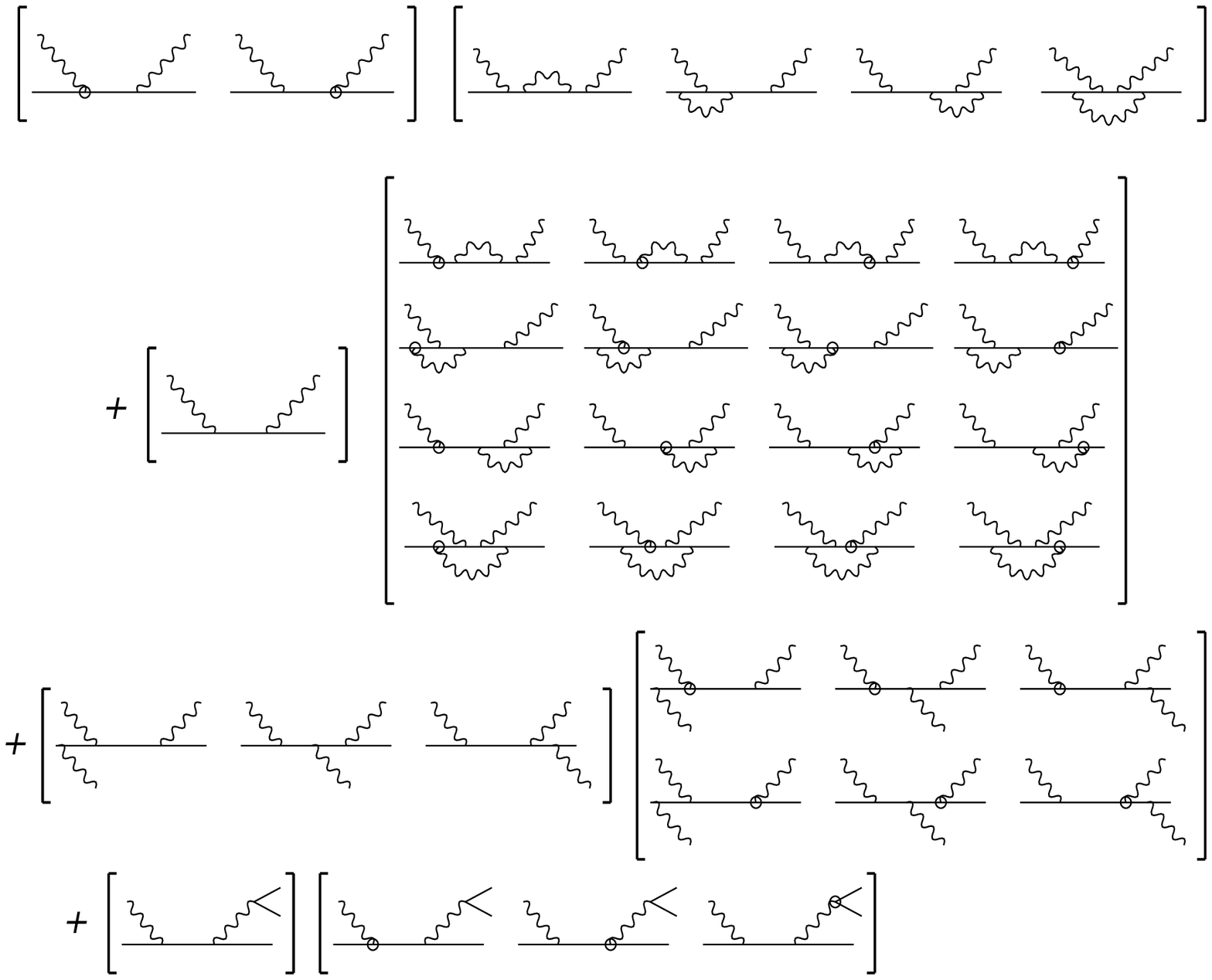}
}
\caption{Interference contributions which give rise to the
$\De\si'^{(3)}$ in QED.
The circled vertex corresponds to the Pauli coupling.}
\figlab{qed2loop}
\end{figure}
\end{document}